# Quantum Imaging of Topologically Unpaired Spin-Polarized Dirac Fermions

Kenjiro K. Gomes,[1,4†] Wonhee Ko,[2,4†] Warren Mar,[3,4†] Yulin Chen,[1,4] Zhi-Xun Shen,[1,2,4] Hari C. Manoharan[1,4*]

[1]*Department of Physics, Stanford University, Stanford, CA 94305, USA*
[2]*Department of Applied Physics, Stanford University, Stanford, CA 94305, USA*
[3]*Department of Electrical Engineering, Stanford University, Stanford, CA 94305, USA*
[4]*Stanford Institute for Materials and Energy Science, SLAC National Accelerator Laboratory, Menlo Park, CA 94025, USA*

[†]These authors contributed equally to this work.
[*]To whom correspondence should be addressed. E-mail: manoharan@stanford.edu

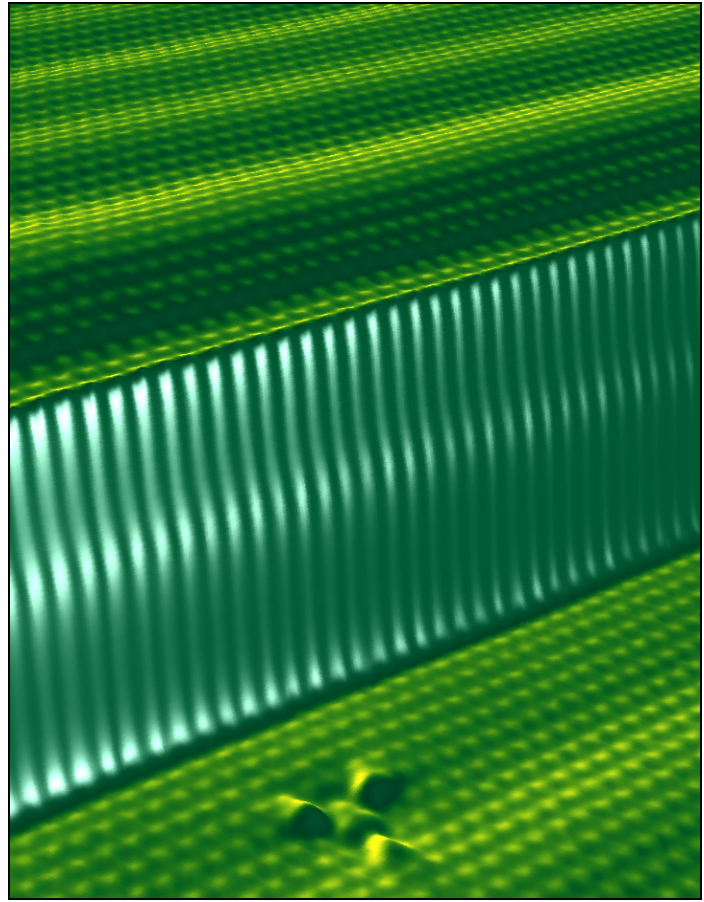

*Electrons on the surface of antimony comprise a "topological metal," mimicking chiral relativistic particles and giving rise to coherent nanoscale spin currents.*

**The motion of a relativistic particle is linked to its spin by the Dirac equation. Remarkably, electrons in two-dimensional materials can mimic such Dirac particles but must always appear in pairs of opposite spin chirality. Using topological ideas of chiral boundary electrons wrapping a three-dimensional crystal with tuned spin-orbit coupling, we show that elusive unpaired Dirac fermions exist on the surface of a topological insulator parent matrix, pure antimony. Using scanning tunneling microscopy and angle-resolved photoemission spectroscopy we image coherent quantum interference between these particles, map their helical spin texture, and ultimately observe a transition to a single Dirac fermion distinguished by backscattering suppression via Berry phase interference. The robust dynamics of these unique spin-polarized carriers envisage future applications in spintronics and topological quantum computation.**

The quest for a condensed-matter analog of chiral relativistic particles has a long theoretical history but only recent experimental success. Typically focused on the symmetries of the two-dimensional (2D) honeycomb lattice, where the two hexagonal sublattices can be mapped to a pseudospin, tight-binding models can approximate the physics of the Dirac equation for quasiparticles at low energies (*1*). These ideas were brought to center stage with the experimental realization of graphene and the discovery of a "relativistic" quantum Hall (QH) effect with a magnetic field applied perpendicular to the monolayer (*2, 3*). While graphene is emerging as a prototype system for studying quantum electrodynamics "in a chip," it is nevertheless subject to a seemingly unbreakable fermion doubling rule (*4*), which requires that all Dirac fermions realized on a time-reversal (TR) invariant 2D lattice must appear in pairs of opposite chirality (leading to the twofold valley degeneracy); moreover, in graphene it is the pseudospin and not real spin that has a well-defined quantum link to quasiparticle momentum via the effective Hamiltonian (preserving the twofold spin degeneracy). Hence in graphene there exist four degenerate copies of a Dirac fermion, each contributing a half conductance quantum ½($e^2$/$h$) to the total observed QH conductance of 2$e^2$/$h$ (*2, 3*), and no net chirality or spin polarization exists.

There are exciting scientific and technological motivations to break these limitations in new quantum materials. On the fundamental side, finding a physical realization for unpaired Dirac fermions would provide for the first time a laboratory to study (2+1)-D space-time relativistic quantum field theories that predict exotic anomalies yet to be observed with real relativistic particles (*1, 5-7*). On the practical side, unpaired Dirac fermions linked to true spin constitute a novel form of relativistic spin-polarized carriers; they open the door to state-of-the-art spintronics applications (*8*) and can be used as a starting ingredient for topological quantum computing platforms (*9*). Fueled by this potential, recent attention has focused on the edges or boundaries of materials possessing embedded topological order. Edge states have a rich history stemming from the physics of the QH effect (*10*), wherein bulk 2D electron conductivity is gapped out by a large magnetic field, and quantized conductance channels form at the one-dimensional (1D) edge of the resulting 2D insulator. Electrons in these states are chiral, propagating only in one direction set by the magnetic field vector, thus flowing without dissipation since no backscattering path exists. However, this chirality is only related to the flow of charge around the edges, and is unrelated to spin, which is frozen out by the large field which breaks TR symmetry. Recently a close sibling of the quantum Hall effect—the quantum spin Hall (QSH) effect (*11-13*)—was identified as a new quantum phase of 2D electrons (*14, 15*). This state, characterized by full TR invariance and no external magnetic field, can be enabled by spin-orbit coupling (SOC) in the bulk that gives rise to inverted band structures: solids typically have conduction band (CB) minima composed of bonding orbitals (characterized by symmetric wavefunctions), and valence band (VB) maxima composed of antibonding orbitals (with antisymmetric wavefunctions); however with sufficiently large SOC, this order can be reversed, embedding a topological "twist" in the underlying electronic states (*13*). Starting with a 2D band structure, this recipe gives a special set of 1D counterpropagating edge states of opposite spin polarization. Distinguished from the chiral variety observed in the QH phase, these states carry a net spin current and are often denoted helical edge states. This helicity of the quasiparticles inhibits backscattering, because excitations with opposite momentum $(+\mathbf{k}, -\mathbf{k})$ also have opposing spins $(\uparrow, \downarrow)$ which cannot mix unless TR symmetry is broken.

The QSH phase in three dimensions (3D) has now entered the scene in a series of rapid developments (*16-21*). It is now appreciated that the 2D and 3D QSH phases are dimensional analogs occurring in an entirely new class of materials—topological insulators—which embed an inverted band structure and realize gapless spin-helical boundary modes (*14, 22*). In the higher-dimensional case, the "edge" of the 3D bulk is a surface comprising a 2D gas of Dirac fermions of definite helical spin polarization. The 2D surface state (SS) itself constitutes a new phase of matter, known as a topological metal (*16, 20*), with an extraordinary property that an odd number of Dirac fermions can be materialized on a given surface, ostensibly circumventing the fermion doubling rule. In fact the pairing still occurs but on opposite surfaces, leading to the conceptual picture that single Dirac fermion species of opposite helicity are "holographically" projected into the top and bottom lower-dimensional surfaces from the higher-dimensional bulk (*23*). Thus in principle, measurements and devices fabricated on one surface only have access to these topologically unpaired Dirac fermions.

Experimental knowledge of this new 2D electronic phase comes largely from angle-resolved photoemission spectroscopy (ARPES) (*21, 24-27*), which allows direct measurement of the underlying band structure limited to energies below the Fermi energy $E_\text{F}$, and the determination of the spin texture and Berry's phase with spin-resolved ARPES (*24, 25, 27*). Direct measurements of the quasiparticle dynamics, quantum transport, and spin-dependent scattering are crucial to future applications and can be addressed with scanning tunneling microscopy (STM), recently applied to non-elemental versions of these systems (*28-30*). In this work we apply STM and scanning tunneling spectroscopy (STS) at 4.2 K to directly image topological Dirac fermions in their simplest elemental incarnation, through spatial mapping of their wavefunctions $\psi(\mathbf{r})$ and coherent quasiparticle interference (QPI), and accurately correlate these real-space (**r**-space) measurements with momentum-space (**k**-space) ARPES experiments we perform at 10 K on the same materials (see SOM Methods). We also perform 1D (via step edges) and 2D (via atomic defects) spin-filtered interferometry on these states to unambiguously visualize the unique spin-textured topological metal ground state.

Any elemental crystal with up to two atoms per primitive unit cell must have an inversion symmetry center (*31*), requiring that $E(\mathbf{k},\uparrow) = E(-\mathbf{k},\uparrow)$. Since TR symmetry requires that $E(\mathbf{k},\uparrow) = E(-\mathbf{k},\downarrow)$—the Kramers degeneracy—spin degeneracy in the bulk is preserved. At a surface termination, however, spatial inversion symmetry is broken allowing $E(\mathbf{k},\uparrow) \neq E(\mathbf{k},\downarrow)$. For example, a strong SOC is capable of lifting the spin degeneracy of normal Shockley SSs (*32*) giving rise to two shifted parabolic subbands in materials such as Au(111) (*33*). Figure 1A, inset, diagrams the fundamental difference between this (so-called topologically trivial) case and the non-trivial situation in topological insulators, where the spin-polarized SSs are a consequence of the special topology of the conduction and valence bands, which can be determined from the parity set of the Bloch wavefunctions of the bulk crystal (*20*). These SSs continuously connect to both the CB and the VB forming a robust band that cannot be gapped out. TR symmetry produces the degeneracy at symmetric points of the Brillouin zone (BZ) and results in the formation of Dirac cones at the band crossings. As long as TR symmetry is preserved in the system, the degeneracy that forms the Dirac points is protected against perturbations. Even if the bands become distorted as shown, any constant-*E* cut between TR-invariant momenta (TRIM), or Kramers points, is guaranteed by topology to cross the SS bands an odd number of times, yielding the goal of unpaired Dirac fermions.

What is the simplest model system to realize such non-trivial topological states? A theoretical method of classification, akin to a topological "periodic table," has identified four $\mathbb{Z}_2$ invariant topological quantum numbers $(\nu_0;\nu_1\nu_2\nu_3)$, where $\nu_i \in \{0,1\}$ are constructed from the even/odd parity sets of the underlying band structure and thus enumerate 16 distinct topological phases (*16-18*). Analogous to $\nu$ serving as a topological classification via an integer Chern number (*10*) for QH states with Hall conductivity $\nu e^2/h$, the quantum number $\nu_0$ distinguishes between two families of topologically trivial $(\nu_0=0)$ and non-trivial $(\nu_0=1)$ QSH states, the latter being robust against disorder and other perturbations. The remaining triplet $(\nu_1\nu_2\nu_3)$ serves as a Miller index of the TRIM. A search through the periodic table reveals promising elemental crystals such as C, Bi, and Au: C in graphene (*11, 13*) or diamond lattices (*16, 20*) can realize 2D and 3D topological QSH phases but only if SOC is artificially added, and Au and Bi as heavy elements have high intrinsic SOC but trivial $\nu_0=0$ topology (*20, 34*). Using this classification and extending the evaluation to specific compounds, it was theoretically predicted (*20*) that the Bi$_{1-x}$Sb$_x$ alloy would be a strong topological insulator $(\nu_0=1)$, and ARPES experiments soon verified this phase in Bi$_{0.9}$Sb$_{0.1}$ by identifying 5 surface band crossings at $E_F$ (*21*). A subsequent search for the origin of topological order in this first generation of topological insulators has centered on pure Sb, since Bi has a trivial (0;000) topology but Sb and a suitably tuned Bi-Sb alloy share in common the (1;111) topological class belonging to the strong topological insulator family (*34*). This provides a route for further experimental investigation: (111) surfaces of pure Sb should exhibit the exotic topological metal phase with SSs composed of odd numbers of Dirac fermions. Methods for identification include projecting the TRIM into the (111) surface Brillouin zone (SBZ) and counting band crossings at $E_F$, a method adopted by recent ARPES measurements combined with spin resolution (*24*). Here we focus on **r**-space imaging of the actual Dirac fermions formed, and directly measure their coherent dynamics by tracking their nanoscale quantum transport governed by charge and spin rules and an embedded Berry's phase unique to a topological metal.

Sb has a finite direct gap throughout the BZ, but is a semimetal because of a negative indirect gap (*35*). Theoretically the topology and the dynamics of the SS depend only on the existence of a direct gap, since the topological classification is based on the parity of bulk bands alone which here give $v_0 = 1$ due to a crucial band inversion at the L point (*20, 34*). We show experimental evidence for this and find that the STM ability to interrogate the SS is not attached to the presence of an indirect gap at nonzero **k**; this yields promise for proposed devices utilizing tunneling contacts to the SS.

A high-resolution STM topograph (Fig. 1A) of the atomically clean Sb(111) surface contains a wealth of information. At the smallest length scales, atomic resolution reveals a clear triangular lattice in which the measured distance between nearest-neighbor atoms is $4.3 \pm 0.05$ Å and the observed atomic step height is $3.7 \pm 0.1$ Å. Both measures are in exceptional agreement with the atomic structure determined by x-ray diffraction at 4.2 K (*36*). Each step observed in STM corresponds to a succession of two atomic layers of the crystal, and steps corresponding to isolated atomic layers are absent, in agreement with the Sb unit cell structure (Fig. 1A, inset). Sb has a rhombohedral crystal structure consisting of two interpenetrating fcc lattices displaced and trigonally distorted along the (111) direction. This results in a natural bilayer pairing of close-packed atomic layers where each atom has 3 nearest neighbors in the adjacent monolayer, and 3 next-nearest neighbors in the more distant plane belonging to the next bilayer. Atoms within a bilayer thus have a honeycomb-like arrangement and are strongly coupled, and the interbilayer coupling is much weaker. With very high spatial resolution, this structure is visible in Fig. 1A via the features within the step edge itself, revealing that the interlayer spacings alternate between $1.5 \pm 0.1$ Å and $2.2 \pm 0.1$ Å, a difference of nearly 50%. This symmetry breaking away from fcc toward honeycomb-like bilayers is a key for realizing the strong topological insulator class (*37*).

Perhaps most striking in Fig. 1A, however, are the clear wave patterns emanating from the step defect that are incommensurate with the atomic structure. At first glance, they look very similar to electronic standing waves observed on noble metals (*38*). As we will show, these oscillations do in fact originate from an Sb(111) SS, but a closer look reveals that the nature of this SS is highly unconventional. For example, the observed **r**-space wavelength $\lambda = 2\pi/q = 34.3$ Å at $E_F$ does not correspond to any wavevector $\mathbf{q} \equiv \mathbf{k}' - \mathbf{k} = 2\mathbf{k}_F$ for the SS band ($\mathbf{k}_F$ denote Fermi wavevectors), as expected for typical Friedel oscillations resulting from quasiparticle scattering between initial and final momentum eigenstates $|\mathbf{k}\rangle$ and $|\mathbf{k}'\rangle$. This can be seen most clearly in ARPES measurements on the same surface (Fig. 1B) registered to **k** within the hexagonal SBZ (Fig. 1B, inset), where the L point projects onto $\overline{M}$, and Γ onto $\overline{\Gamma}$. The SS bands emanate from a Dirac point ($E_D$) 230 meV below $E_F$ at $\overline{\Gamma}$. These bands can be viewed as a typical double-cone Dirac structure, but with the lower cone distorted and folded upward. The top half of the Dirac cone forms an electron pocket that extends to the CB. The lower half of the cone forms an unusual hole band that, even though it is ultimately connected to the VB, bends back above the Dirac point for finite regions in **k**-space. The hole band extends up to 120 meV below $E_F$ along the $\overline{\Gamma} - \overline{K}$ direction and up to 220 meV above $E_F$ along the $\overline{\Gamma} - \overline{M}$ direction (Fig. 1B) before turning back to the VB. The SSs thus indicate a well-resolved $v_0 = 1$ Dirac structure, do not overlap bulk states, and are much simpler than the $Bi_{1-x}Sb_x$ system (*21, 28*).

This electronic structure creates a unique opportunity to study the topological SSs in 3 different regimes (Fig. 1, B and C): (I) from 230 meV to 120 meV below $E_F$ the SSs form an electron pocket around $\bar{\Gamma}$ surrounded by a circumferential holes at larger **k**; (II) from 120 meV below $E_F$ to 220 meV above $E_F$ there is a $\bar{\Gamma}$ electron pocket surrounded by 6 hole pockets centered along the $\bar{\Gamma} - \bar{M}$ directions; (III) above 220 meV and up to the CB [located $> 400$ meV above $E_F$ (*35*)] there is only a single electron pocket centered at $\bar{\Gamma}$. Since ARPES only measures states below $E_F$, we have filled in the band structure above $E_F$ using theory fitted to the data (*35*). As STM has no energy window limitation, we can correlate the essential features of the overall band structure (Fig. 1B) to the spatially averaged tunneling spectrum (Fig. 1C). In STS, sample bias $V$ drives tunnel current $I$ detected at the tip location **r**, and $g(\mathbf{r}, E) \equiv dI/dV(\mathbf{r}, E)$ is the measured differential tunneling conductance ($E = eV = 0$ at $E_F$), proportional to the local density of states (LDOS). A spatial average $\langle g \rangle = \langle g(\mathbf{r}, E) \rangle$ is shown for useful comparison to the DOS and because, as we show later, the LDOS contains other detailed information to analyze. The bottom of the SS band at $\bar{\Gamma}$ and $E_D$ results in a step in the DOS clearly seen at $V = -230$ mV (Fig. 1, B and C, lower arrow), and the extrema in the band structure along the $\bar{\Gamma} - \bar{K}$ and $\bar{\Gamma} - \bar{M}$ directions correlate with peaks in $\langle g \rangle$ at $V = -120$ mV and $V = 220$ mV (Fig. 1, B and C, upper arrows).

On the cleanest surfaces and close to $E_F$, robust **r**-space quantum oscillations appear as waves as high as ~200 mÅ in electronic topography, and exhibit phase coherence over the largest images acquired (~1000 Å). This is a noticeably larger perturbation than observed on noble metals, but surprisingly no STM report of the Sb(111) SS exists (*28-30*). We turn to a detailed characterization of it. Fig. 2A shows a sequence of $g(\mathbf{r}, E)$ measurements acquired at equally spaced distances $r = d$ along a line perpendicular to a step edge (left topograph in Fig. 2A). The ripples observed in the spectra are pronounced in a $V$ window between $\sim \pm 200$ mV, in which clear dispersive behavior is manifested. The wavelength of the oscillation in the LDOS is best observed in the normalized conductance $g(d, eV)/\langle g \rangle$, plotted in Fig. 2B for $V$ between $\pm 100$ mV and $d$ increasing to 250 Å from the step edge.

The LDOS oscillations are a result of QPI and the formation of standing waves from the electrons scattering at the 1D step defect (*29, 30*). In order to extract the wavevector $q$ of these oscillations, we employ a simple strong-scattering step edge model (*38*) that is still valid in the presence of SOC (*39*): $g(d, eV) = \langle g \rangle [1 - J_0(qd)]$, where $J_0$ is the zeroth-order Bessel function of the first kind. This model with $q$ as a single fitting parameter produces excellent agreement with the data, as exemplified by the measured and fit $g(d, E_F)$ displayed in Fig. 2C.

We compare the $q(E)$ dispersion measured from the **r**-space STS data to all possible changes in momentum $k' - k$ due to backscattering extracted from the ARPES band structure (Fig. 2D). The step edge facet is $(1\bar{1}2)$ which runs along the close-packed atom rows as evident in the Fig. 1A topograph; this leads to backscattering along the $\bar{\Gamma} - \bar{M}$ direction. In region II studied here (Fig. 1, B and C), therefore, there are 3 possible backscattering scenarios: quasiparticles can scatter within the electron pocket, scatter between opposing hole pockets, or scatter between the electron and the opposite hole pocket. The precision of the ARPES and STS measurements to determine $q$ allows us to pinpoint that we only observe scattering across counterpropagating electron and hole pockets. This result is one of the central observations of this work, as it is a testament of the spin texture present in the SS (see Fig. 2D, inset). As states with exact opposite momentum also have opposite spins, backscattering within the electron pocket or between opposing hole pockets should be forbidden by spin selection rules, as the step edge potential is spinless and does not break TR symmetry. As we overlap the STS results on the 3 possible ARPES backscattering scenarios (2 involving spin flips, and one in which spin is conserved), we see very graphically that the coherent dynamics of the SS involve only spin-polarized carriers that maintain spin coherence throughout the scattering process. Indeed these dynamics are so robust that the Dirac particles break momentum conservation in order to preserve their spin.

We extend our analysis to the 2D case, where the Dirac particles scattering from lattice defects and surface impurities (Fig. 3, inset) form standing wave patterns along multiple directions (*28, 30*). We map the complete LDOS by measuring $g(\mathbf{r}, E)$, acquiring *dI/dV* vs *V* at each pixel of an atomic resolution scan over a 450 Å × 450 Å region. Slices through this dataset at representative *V* are shown in Fig. 3. The observed electronic wave patterns from QPI are not isotropic but instead are dominated by scattering along the $\overline{\Gamma} - \overline{M}$ direction (dispersing sixfold-symmetric patterns are visible throughout Fig. 3, A to C). Notably, the dispersion of the standing wave pattern is only observed up to ~200 mV, near the top of region II in Fig. 1 and consistent with Fig. 2A. As *V* is increased above this point there is a striking transition to strongly suppressed QPI, and maps (e.g. Fig. 3D) become homogenous with no well-defined modulation patterns. We show below this is due to a crossover to a single spin-polarized Dirac cone that is topologically protected.

The characterization of the patterns found in the $g(\mathbf{r}, E)$ maps is achieved through the combination of a Fourier-transform analysis (FT-STS) and the determination of possible scattering vectors **q** from the ARPES constant-*E* contours of the band structure. The Fermi surface (within region II of Fig. 1) is composed of a central electron pocket around the $\overline{\Gamma}$ point, surrounded by 6 hole pockets along $\overline{\Gamma} - \overline{M}$ (Fig. 4A). The ARPES intensity can be interpreted as a map of the quasiparticle spectral function $A(\mathbf{k}, E)$ giving the probability of adding or removing an electron at a given **k** and *E*. In STM, QPI stems from quasiparticles scattering elastically from potential *U* at rate $\Gamma_{\mathbf{k} \to \mathbf{k'}} = (2\pi / \hbar) |\langle \mathbf{k'} | U | \mathbf{k} \rangle|^2 \delta(E_{\mathbf{k'}} - E_{\mathbf{k}})$ using Fermi's golden rule. The total probability for a scattering event with wavevector shift **q** at a given *E* is

$$\Gamma(\mathbf{q}, E) = \sum_{\mathbf{k},\mathbf{k'}} \Gamma_{\mathbf{k} \to \mathbf{k'}} = \frac{2\pi}{\hbar} \sum_{\mathbf{k},\mathbf{k'}} |\langle \mathbf{k'} | U | \mathbf{k} \rangle|^2 \delta(E_{\mathbf{k'}} - E) \delta(E_{\mathbf{k}} - E)$$
$$\approx \frac{2\pi}{\hbar} |\langle \mathbf{k} + \mathbf{q} | U | \mathbf{k} \rangle|^2 \gamma(\mathbf{q}, E)$$
(1)

where $\gamma(\mathbf{q}, E) = \sum_{\mathbf{k}} \delta(E_{\mathbf{k}+\mathbf{q}} - E) \delta(E_{\mathbf{k}} - E)$ is the joint DOS (JDOS) which can be directly related to the autocorrelation (AC) of the ARPES spectral function,

$$\gamma(\mathbf{q},E) = \int A(\mathbf{k}+\mathbf{q},E) A(\mathbf{k},E) d\mathbf{k}. \qquad (2)$$

Since STS measures $g(\mathbf{r},E) \propto \text{LDOS}(\mathbf{r},E) \propto \sum_{\mathbf{k}} |\psi_{\mathbf{k}}(\mathbf{r})|^2 \delta(E_{\mathbf{k}} - E)$, FT-STS measures $\mathcal{F}[g(\mathbf{r},E)] = g(\mathbf{q},E)$ which contains all the momentum contributions of the JDOS $\gamma(\mathbf{q},E)$ through QPI. This equivalence has been exploited to bridge FT-STS and ARPES data in high-temperature superconductors (*40-42*) and apparently works because the approximation in Eq. (1) is typically valid, i.e. the matrix element $|\langle \mathbf{k}'|U|\mathbf{k}\rangle|^2$ can be removed from the sum. Here we show that this assumption fails spectacularly because of the spin texture of the underlying Dirac quasiparticles, but a modified *spin-polarized* AC (SPAC) procedure can faithfully reconstruct all quantum degrees of freedom even without the need for direct spin measurement (*28*).

We first evaluate the main QPI contributions in the absence of spin polarization by constructing a model $A(\mathbf{k},E)$ from the ARPES data (Fig. 4A) and computing the AC $\gamma(\mathbf{q},E)$ using Eq. (2) (Fig. 4B). The hexagonal shape of the central electron pocket gives rise to nesting vectors that manifest as the most intense $\mathbf{q}$ vectors in the $\gamma$-map (Fig. 4B). These $\mathbf{q}$ vectors, parallel to the $\overline{\Gamma}-\overline{K}$ directions, are typical $2\mathbf{k}_F$ backscattering vectors seen in normal metals; for the topological metal, however, they connect opposite sides of the electron pocket which must have opposite spin and therefore need to be filtered out from the calculations. This can be seen vividly by comparing $\gamma(\mathbf{q},E)$ to the experimental $g(\mathbf{q},E)$ FT-STS data (Fig. 4D), which looks profoundly different and is characterized by a distinct absence of $2\mathbf{k}_F$ QPI. We find that the main contributions in the *g*-maps arise from QPI along the $\overline{\Gamma}-\overline{M}$ directions corresponding to backscattering between the edge of the hole pockets and the opposing end of the electron pocket. This is in fact the singular $\mathbf{q}$ vector highlighted in Fig. 2D which conserves spin.

We now construct a general procedure to extract the underlying spin texture from the STM data. The essential ingredient is the intrinsic link between a Dirac fermion's momentum $\mathbf{k}$ and its spin $\mathbf{s}_{\mathbf{k}}$. To accommodate we do not make the approximation in Eq. (1) but keep the spin part of the scattering matrix element in the JDOS, leading to the SPAC function

$$\gamma_{SP}(\mathbf{q},E) = \int |\langle \mathbf{s}_{\mathbf{k}+\mathbf{q}}|\mathbf{s}_{\mathbf{k}}\rangle|^2 A(\mathbf{k}+\mathbf{q},E) A(\mathbf{k},E) d\mathbf{k} \qquad (3)$$

where $U$ is assumed to be TR invariant and thus factors out of the spin matrix element. With SOC, the spin texture of the SS can be represented by the spinor $\mathbf{s}_{\mathbf{k}} = [1, \pm i \exp(i\theta_{\mathbf{k}})]$, where $\theta_{\mathbf{k}} = \tan^{-1} k_y/k_x$ is the angle of propagation in $\mathbf{k}$-space relative to the *x*-axis and $\pm$ refers to the spin helicity of the Dirac fermion. The inclusion of spin therefore builds in a non-trivial $|\langle \mathbf{s}_{\mathbf{k}'}|\mathbf{s}_{\mathbf{k}}\rangle|^2 = \cos^2[(\theta_{\mathbf{k}'}-\theta_{\mathbf{k}})/2]$ dependence for a Dirac fermion scattering from $\mathbf{k}$ to $\mathbf{k}'$, a direct consequence of its spin rotation by $\Delta\theta = \theta_{\mathbf{k}'} - \theta_{\mathbf{k}}$. The elaboration and application of spin rules in topological insulator QPI was first applied recently to Bi$_{0.92}$Sb$_{0.08}$ using STM and spin-ARPES (*28*). By calculating the SPAC with the spin texture (Fig. 4C), which includes electron and hole pockets of opposite helicity, we obtain remarkable agreement between $\gamma_{SP}(\mathbf{q},E)$ (Fig. 4C) and the experimental $g(\mathbf{q},E)$ (Fig. 4D). We note that this method is even more general than that presented above: we find that $\theta_{\mathbf{k}}$ can be used as a set of fitted spin orientations (i.e. without assuming any specific relation between $\mathbf{k}$ and $\theta_{\mathbf{k}}$) allowing the underlying spin texture to be extracted by iterative comparison between $\gamma_{SP}(\mathbf{q}, E)$ and the STM data $g(\mathbf{q}, E)$ at any $E$. This Fourier-space spin extraction procedure is based on a real-space quantum phase extraction procedure in nanostructures (*43*). Surprisingly this allows a detailed spin measurement without direct spin sensitivity in the STM or ARPES. In this measurement the hole pocket can be viewed as a spin-polarized injector of carriers and the electron pocket a spin-filtered detector, with the spin textures ultimately being deciphered via QPI Fourier amplitude.

The full $E$ dependence of the FT-STS data is shown in Fig. 4E, revealing the spin-polarized dispersion extrapolating back to the Dirac point. Using the SPAC procedures above and combining the ARPES and FT-STS data, we can fully reconstruct the SS topology and spin texture purely from experimental data in regions I and II (Fig. 4E, right). By moving into region III, the STS data show a prominent phenomenon in an $E$ window that cannot be probed by ARPES. The **q** vector representing scattering events between the hole and electron pockets vanishes for voltages above ~200 mV (Fig. 4E). This corresponds to the $E$ at which the hole pocket closes (see Fig. 1, B and C) leaving only the electron pocket—a single unpaired Dirac fermion. In this regime, the helicity of the Dirac cone forbids backscattering and results in the absence of QPI. The only self-consistent fit to the spin texture that results from the SPAC procedure is shown in Fig. 4E, right, which demonstrates explicitly that this regime is distinguished by a single protected Dirac species that embeds a net helicity and accompanying $\pi$ Berry's phase upon one $\Delta\theta = 2\pi$ circuit in **k**-space (*16, 18, 20, 24, 25, 27*). The **r**-space quenching of localized QPI (Fig. 3) gives a vivid picture of the onset of antilocalization in this regime, resulting from destructive interference of time-reversed paths in backscattering trajectories due to the topological Berry's phase. The Dirac fermions can now flow with high mobility and carry a spin current around the same defects that caused backscattering at lower $V$.

We note in closing that one limitation of the SPAC and FT-STS techniques presented here is that—while remarkably yielding all physically important spin textures and proving the presence and dynamics of a topological metal—they formally determine the absolute spin polarization up to an overall global phase equivalent to a gauge transformation. However, it suffices to measure the absolute spin polarization at a single point in **k**-space and then the entire spin structure is completely determined (see SOM and Fig. S1). This determines the net helicity direction displayed in Fig. 4E, which corresponds to left-handed chirality for the single Dirac species observed. This is consistent with the electron pockets measured by spin-resolved ARPES up to $E_F$, but which notably cannot probe the transition to the single Dirac fermion in Sb (*24*). Single Dirac cones can be realized with more complex crystals and doping (*25-27, 29, 30*), but the fusion of STM, QPI, and the SPAC-ARPES method here reveals that unpaired and single spin-polarized Dirac fermions may be far more accessible than previously imagined. In a relatively simple material such as Sb, topology already separates the Fermi surface to enclose the TRIM an odd number of times and allows for spin-interferometry of the SSs, and then a straightforward $V$ change gives access to an elusive single species of Dirac particle with helical spin texture and non-trivial Berry's phase. As such carriers form the foundation for efforts to realize relativistic electrodynamics and quantum effects not seen with other matter—including axion fields from quark physics (*19*), induced magnetic monopoles (*44*), and Majorana fermions (*45, 46*)—the observed requisite spin protection on an elemental crystal expands the realm of possibilities for future "Dirac devices."

**References and Notes**

1. G. W. Semenoff, *Physical Review Letters* **53**, 2449-2452 (1984).
2. K. S. Novoselov *et al.*, *Nature* **438**, 197-200 (2005).
3. Y. Zhang, Y.-W. Tan, H. L. Stormer, P. Kim, *Nature* **438**, 201-204 (2005).
4. H. B. Nielsen, M. Ninomiya, *Physics Letters B* **105**, 219-223 (1981).
5. E. Fradkin, E. Dagotto, D. Boyanovsky, *Physical Review Letters* **57**, 2967-2970 (1986).
6. F. D. M. Haldane, *Physical Review Letters* **61**, 2015-2018 (1988).
7. F. Wilczek, *Physical Review Letters* **58**, 1799-1802 (1987).
8. D. Awschalom, N. Samarth, *Physics* **2**, 50 (2009).
9. A. Y. Kitaev, *Annals of Physics* **303**, 2-30 (2003).
10. *Perspectives in Quantum Hall Effects: Novel Quantum Liquids in Low-Dimensional Semiconductor Structures*. S. Das Sarma, A. Pinczuk, Eds. (Wiley, New York, 1997).
11. C. L. Kane, E. J. Mele, *Physical Review Letters* **95**, 226801 (2005).
12. B. A. Bernevig, S. C. Zhang, *Physical Review Letters* **96**, 106802 (2006).
13. C. L. Kane, E. J. Mele, *Physical Review Letters* **95**, 146802 (2005).
14. B. A. Bernevig, T. L. Hughes, S. C. Zhang, *Science* **314**, 1757-1761 (2006).
15. M. Konig *et al.*, *Science* **318**, 766-770 (2007).
16. L. Fu, C. L. Kane, E. J. Mele, *Physical Review Letters* **98**, 106803 (2007).
17. J. E. Moore, L. Balents, *Physical Review B* **75**, 121306(R) (2007).
18. R. Roy, *Physical Review B* **79**, 195322 (2009).
19. X.-L. Qi, T. L. Hughes, S.-C. Zhang, *Physical Review B* **78**, 195424 (2008).
20. L. Fu, C. L. Kane, *Physical Review B* **76**, 045302 (2007).
21. D. Hsieh *et al.*, *Nature* **452**, 970-974 (2008).
22. H. Zhang *et al.*, *Nature Physics* **5**, 438-442 (2009).
23. C. Wu, B. A. Bernevig, S.-C. Zhang, *Physical Review Letters* **96**, 106401 (2006).
24. D. Hsieh *et al.*, *Science* **323**, 919-922 (2009).
25. Y. Xia *et al.*, *Nature Physics* **5**, 398-402 (2009).
26. Y. L. Chen *et al.*, *Science* **325**, 178-181 (2009).
27. D. Hsieh *et al.*, *Nature* **460**, 1101-1105 (2009).
28. P. Roushan *et al.*, *Nature* **460**, 1106-1109 (2009).
29. Z. Alpichshev *et al.*, arXiv:0908.0371 (2009).
30. T. Zhang *et al.*, arXiv:0908.4136 (2009).
31. R. Zallen, *Physical Review* **173**, 824-832 (1968).
32. W. Shockley, *Physical Review* **56**, 317-323 (1939).
33. S. LaShell, B. A. McDougall, E. Jensen, *Physical Review Letters* **77**, 3419-3422 (1996).
34. J. C. Y. Teo, L. Fu, C. L. Kane, *Physical Review B* **78**, 045426 (2008).
35. Y. Liu, R. E. Allen, *Physical Review B* **52**, 1566-1577 (1995).
36. C. S. Barrett, P. Cucka, K. Haefner, *Acta Crystallographica* **16**, 451-453 (1963).
37. S. Murakami, *Physical Review Letters* **97**, 236805 (2006).
38. M. F. Crommie, C. P. Lutz, D. M. Eigler, *Nature* **363**, 524-527 (1993).
39. J. D. Walls, E. J. Heller, *Nano Letters* **7**, 3377-3382 (2007).
40. R. S. Markiewicz, *Physical Review B* **69**, 214517 (2004).
41. K. McElroy *et al.*, *Physical Review Letters* **96**, 067005 (2006).
42. U. Chatterjee *et al.*, *Physical Review Letters* **96**, 107006 (2006).
43. C. R. Moon *et al.*, *Science* **319**, 782-787 (2008).
44. X.-L. Qi, R. Li, J. Zang, S.-C. Zhang, *Science* **323**, 1184-1187 (2009).
45. L. Fu, C. L. Kane, *Physical Review Letters* **102**, 216403 (2009).
46. A. R. Akhmerov, J. Nilsson, C. W. J. Beenakker, *Physical Review Letters* **102**, 216404 (2009).

**Acknowledgements**

This work was supported by the Department of Energy, Office of Basic Energy Sciences, Division of Materials Sciences and Engineering, under contract DE-AC02-76SF00515. Nanoprobe development and fabrication were supported by the National Science Foundation. We thank H. Fertig, T. Hughes, A. Ludwig, X.-L. Qi, R. Roy, K. Yang, S.-C. Zhang for discussions, and C. Moon and L. Mattos for technical assistance.

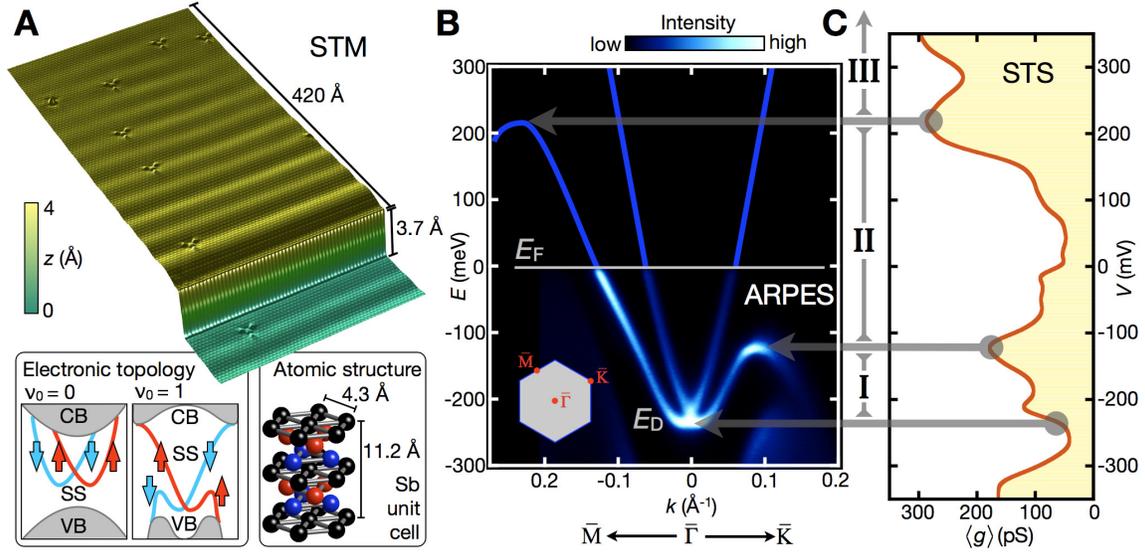

**Fig. 1.** Atomic and electronic structure of Sb(111). (**A**) STM topography of Sb(111) displays both the atomic corrugation and robust electronic QPI oscillations from the SS (230 Å by 500 Å, $V$ = 10 mV, $I$ = 100 pA). The localized 3-fold symmetric features are subsurface defects. Right inset: Sb unit cell showing ABCABC… stacking (black, red, blue triangular lattices) with AB..CA..BC.. bilayer pairing, also visible in the step edge topography. Left inset: Distinct topologies for SSs with SOC; as opposed to the trivial case ($v_0$=0), the Sb classification ($v_0$=1) produces unpaired Dirac fermions. (**B**) ARPES spectrum featuring the SS dispersion along a $\overline{M}-\overline{\Gamma}-\overline{K}$ cut in **k**-space. The Dirac point ($E_D$) is 230 meV below $E_F$. Energies above $E_F$ show band structure from theory using STS and ARPES observations as reference. The central electron pocket is nearly isotropic with $v_F$ ranging from $(7.4\pm0.5)\times10^5$ to $(8.1\pm0.5)\times10^5$ m/s and becomes a single Dirac cone above 220 meV. (**C**) The differential tunneling conductance spectrum $g(E)$, spatially averaged over a 270 Å line, links to features in the ARPES spectrum within the same $E$ window. The separate hole pockets have $v_F = (3.1\pm0.2)\times10^5$ m/s. The bottom of the SS band produces a step in the tunneling spectrum, and other points where $dE/dk = 0$ produce resonances. These features delineate 3 regimes (I, II, III) for Dirac particle interferometry discussed in the text. The complete series of spectra composing this average is shown in Fig. 2A ($I$ = 200 pA, $V$ = 500 mV, $V_{mod}$ = 5 mV rms).

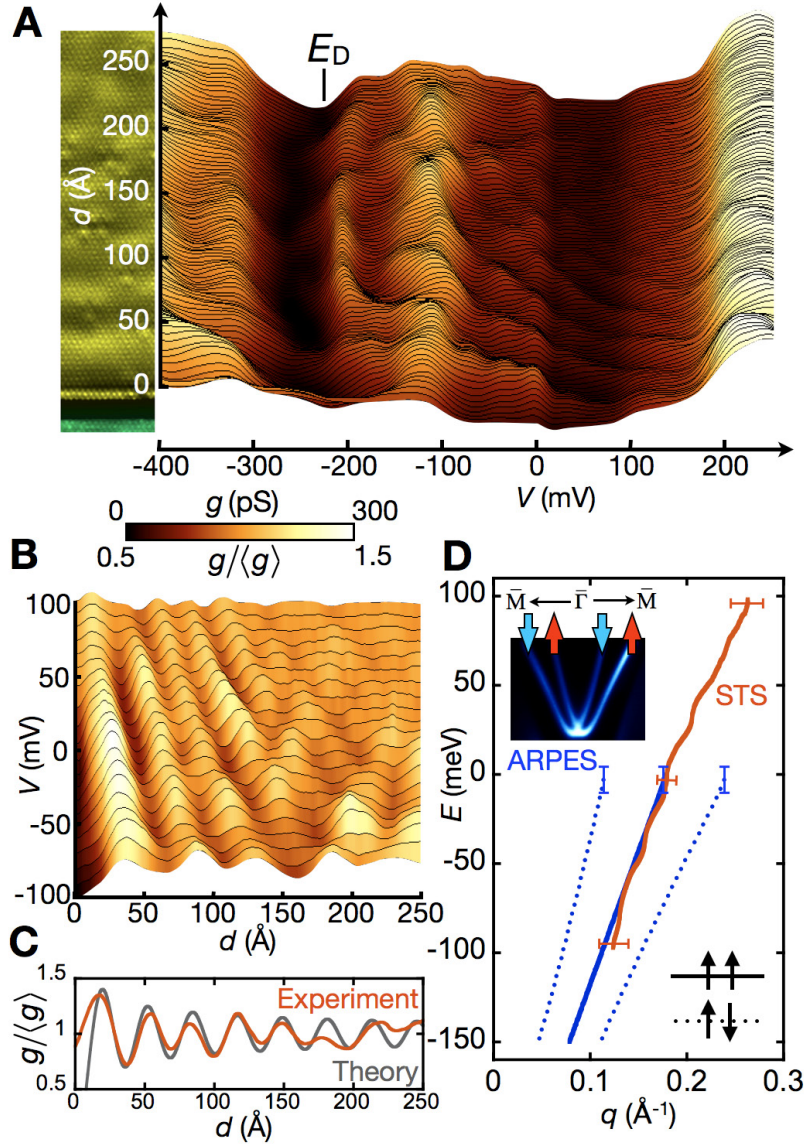

**Fig. 2.** 1D electronic dispersion of the topological metal. (**A**) A series of tunneling spectra $g(d,E)$ measured as a function of distance $d$ from a step edge (left axis topography) shows dispersing spatial modulations. (**B**) Conductance oscillations normalized to the spatial average in a central $E$ window. (**C**) The spatial dependence of the normalized conductance at $E_F$ (orange) is plotted with a theoretical fit (gray). (**D**) The spectrum of the extracted wavenumbers $q$ from the STS oscillations (orange) is compared with all possible backscattering $q$ values (blue) extracted from the ARPES band structure shown in the inset. The ARPES solid line, which precisely overlaps the STM data, is the only spin-conserving backscattering $q$, while spin-flip events (dotted lines) are excluded based on STM observations. All lines shown extrapolate back to $E_D$. Error bars in the ARPES data represent the instrument $E$ resolution (8 meV) and the STM error bar is the average deviation (~0.01 Å$^{-1}$) for different fitting criteria used. The spins lie in the surface plane but are shown as up and down arrows for clarity, with the convention that spin up denotes spin parallel to the $+k_y$ direction.

– 12 –

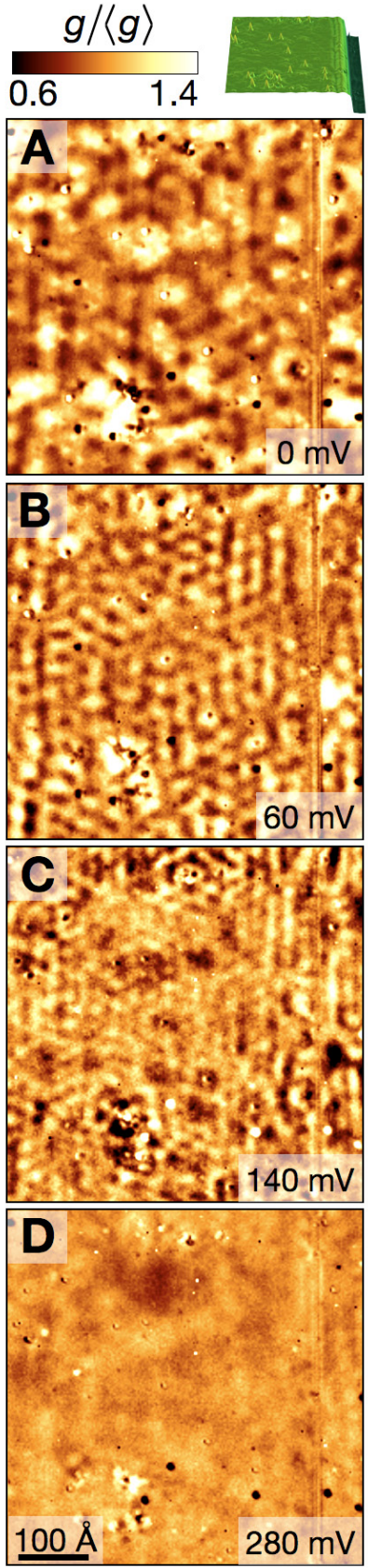

**Fig. 3**. 2D visualization of the topological SS standing waves. (**A-D**) Differential conductance maps $g(\mathbf{r},E)$ measured at each pixel location **r** over a 450 Å by 450 Å area, shown at $V$ slices 0 mV (A), 60 mV (B), 140 mV (C), and 280 mV (D). The area topography is shown in top inset. Each image color scale is normalized by the spatially averaged $g$ for best direct comparison. From (A) to (C), modulations are observed displaying a dispersing wavelength but are absent in the map taken at 280 mV (D), which is uniform without any clear pattern due to a transition to a single protected Dirac fermion species and accompanying antilocalization from Berry phase interference. (Topography: $I$ = 200 pA, $V$ = 100 mV; maps: $I$ = 200 pA, $V$ = 300 mV, $V_{mod}$ = 5 mV rms).

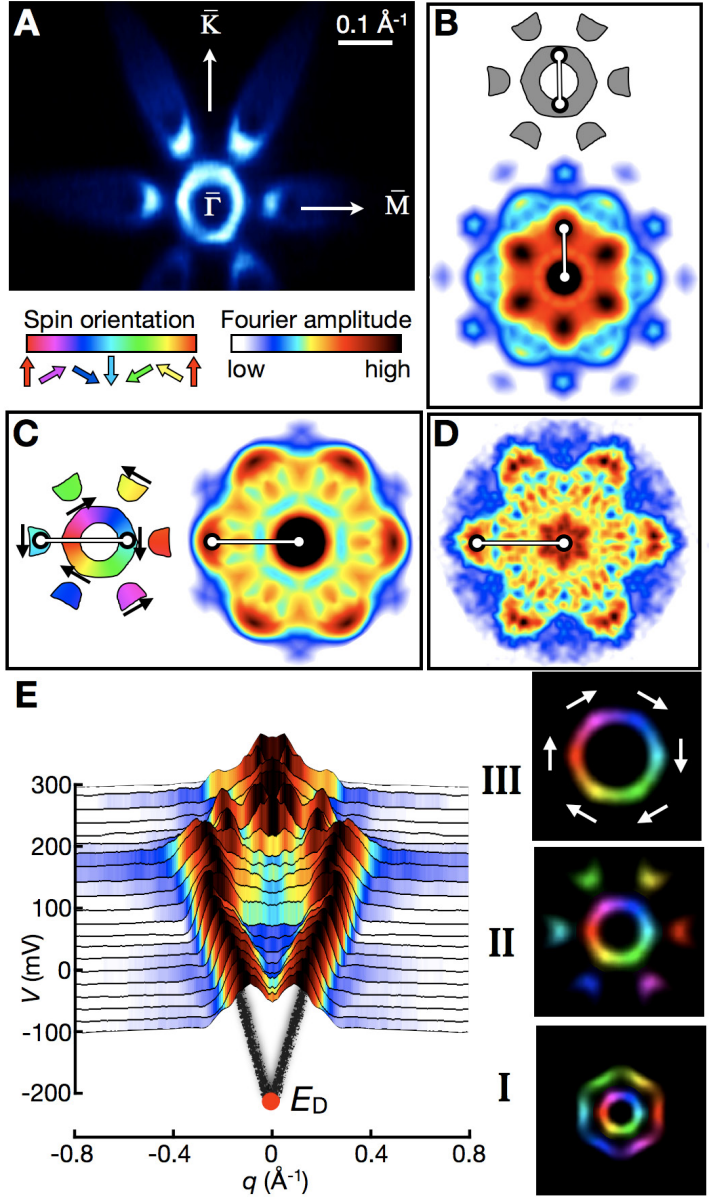

**Fig. 4.** The topological metal spin texture and **k**-space mapping. (**A**) ARPES measurement of the Fermi surface presents a hexagonal central electron pocket surrounded by 6 hole pockets. The faint leaf-shaped extensions of the hole pockets are part of the bulk VB. The FT of the STS data can be related to the AC of constant-$E$ contours of the ARPES spectral function. (**B**) Based on the ARPES measurement, we model the SS constant-$E$ contour (top) to calculate the AC $\gamma(\mathbf{q},E)$ map (bottom). Highest Fourier amplitude appears along nesting directions parallel to $\overline{\Gamma}-\overline{K}$ (vertical white vector). (**C**) We add the spin texture (colors and black arrows) to the electronic states (left) and obtain a new spin-polarized AC (right). Now the highest contribution arises along $\overline{\Gamma}-\overline{M}$ directions from the scattering of Dirac fermions with the same spin (horizontal white vector). (**D**) The FT-STS map $g(\mathbf{q},E)$ displays close correlation to the calculated SPAC map with spin textures in (C), with the same dominant **q** connecting electron and hole pockets along $\overline{\Gamma}-\overline{M}$. The image shown is the FT of the $g(\mathbf{r},E)$ map measured at $V = 60$ mV displayed in Fig. 3B. (**E**) Series of cuts along the $\overline{\Gamma}-\overline{M}$ direction in the FT-STS $g(\mathbf{q},E)$ maps measured from $V = -100$ mV to 300 mV. The observed peaks disperse consistent with the 1D measurement in Fig. 2 and extrapolate to the $E_D$ observed in ARPES (Fig. 1). Right: extracted spin textures combined with constant-$E$ contours in the 3 regimes demarcated in Fig. 1. The images in (I) and (II) are symmetrized versions of the ARPES data measured at −100 meV and 0 meV wrapped with the spin texture extracted from the STM measurements. Image in (III) is an extrapolation of the electron pocket found in (II) wrapped with the self-consistent spin texture from STM. The abrupt change observed at $V \sim 200$ mV in the Fourier amplitude marks the transition to a single unpaired Dirac fermion with left-handed spin helicity.